\documentclass[preprint, aps, 12pt, a4]{revtex4}

\usepackage{graphicx}
\usepackage{dcolumn}
\usepackage{bm}
\usepackage{amsmath}
\usepackage{epsf}
\usepackage{epsfig}
\usepackage{amssymb}
\usepackage{color}
\usepackage{hyperref} 

\begin{document}
\title{ Coulomb Wave Function DVR: Application to
Atomic Systems in Strong Laser Fields\footnote{This manuscript is
prepared for the Journal of Chemical Physics in the Section of
``Theoretical Methods and Algorithms ''.}}

\author{Liang-You Peng and Anthony F. Starace}

\affiliation{Department of Physics and Astronomy, The University of Nebraska-Lincoln,
Nebraska 68588-0111, USA}

\date{\today}

\begin{abstract}
We present an efficient and accurate grid method for solving the
time-dependent Schr\"{o}dinger equation  of atomic systems
interacting with  intense laser pulses. As usual, the angular part
of the wave function is expanded in terms of spherical harmonics.
Instead of the usual finite difference (FD) scheme, the radial
coordinate is discretized using the discrete variable representation
 which is constructed from the Coulomb wave function. For
an accurate description of the ionization dynamics of atomic
systems, the Coulomb wave function discrete variable representation
(CWDVR) method needs 3-10 times less grid points than the FD method.
The resultant grid points of CWDVR distribute unevenly so that one
has finer grid near the origin and coarser one at  larger distances.
The other important advantage of the CWDVR method is that it treats
the Coulomb singularity  accurately and gives a good representation
of continuum wave functions. The time propagation of the wave
function is implemented using the well-known Arnoldi method. As
examples, the present method is applied to the multiphoton
ionization of both H and H$^-$ in intense laser fields. Short-time
excitation and ionization dynamics of H by static electric fields is
also investigated. For a wide range of photon energies and laser
intensities, ionization rates calculated  using this method are in
excellent agreement with those from other theoretical calculations.
\end{abstract}

\maketitle

%
%
%
%

\section{\label{sec0}Introduction}

With the fast advance of modern laser technologies, lasers of
various frequencies at different intensities are routinely available
in many laboratories~\cite{Brab00}.   Studies of the highly
nonlinear interaction of  matters with strong laser pulses have
revealed many interesting physical and chemical
processes~\cite{Prot97}. New technologies based on these physical
principles are under fast development and new frontiers of sciences
are being opened~\cite{Agos04}. Newly developed light sources have
for the first time enabled physicists and chemists to trace and
image the electronic motion within atoms and molecules on attosecond
timescale~\cite{Fohl05}.

However, multiple reactional paths and  many-body nature inherent to
these highly nonlinear and ultrafast processes contribute to the
complexities of theoretical interpretations to  the experimental
observations. High  intensities of the applied lasers make any
perturbation theories no longer applicable. It is necessary to treat
the Coulomb interaction and the interactions from the laser fields
on equal footing. Many theoretical methods have been designed to
describe different kinds of phenomena. To mention just a few, these
methods range from strong field approximation~\cite{Ammo86},
intense-field many-body $S$-matrix theory~\cite{Beck05}, $R$-matrix
Floquet method~\cite{Burk97}, the generalized Floquet
theory~\cite{Chu04}, time-dependent density functional
theory~\cite{Marq04} and numerical integration of the time-dependent
Schr\"odinger equation~\cite{Kula87,Bauer05, Smyt98,Peng04}.
Compared to other methods, the direct solution of the time-dependent
Schr\"odinger equation (TDSE) has proved to be very versatile and
fruitful   in explaining and predicting many experimental
measurements for a wide range of laser parameters. Especially when
the laser pulse length approaches the few-cycle or sub-femtosecond
regime, numerical solution of TDSE becomes even more appropriate and
efficient.

Nevertheless, the accurate integration of the multi-dimensional TDSE
is   very computationally demanding. Even the most powerful
supercomputer nowadays has only  made the {\it ab initio}
integration of the two-electron atom possible~\cite{Meha05}.
Therefore, many approximate theoretical models such as
reduced-dimension models and soft Coulomb potential models have
played an important role in understanding some of the important
physical mechanisms underlying the strong field
phenomena~\cite{Qu91}. On the other hand, it is questionable to use
these models to simulate the realistic experiments {\it
quantitatively}. For example, it was pointed out that the dynamical
motion in three dimensions is not merely a trivial extension of what
happens in two dimensions~\cite{Milc97}. It was also shown that the
physics of model systems using the soft Coulomb potential is very
sensitive to the softening parameters~\cite{Stee03}.

Therefore, in order to produce {\it quantitatively} correct results,
it is necessary to solve TDSE in its full dimensions and use the
real Coulomb potential with its singularity treated
properly~\cite{Kono97}. This is especially crucial when the
singularity plays an important role in the problem at
hand~\cite{Milc97, Comt05}. It is the purpose of the present paper
to present an accurate and efficient method to solve the TDSE for
the  multiphoton ionization of  hydrogenic atom. Unlike the usual
finite difference (FD) discretization of the radial
coordinate~\cite{Kula87,Bauer05}, the present method discretizes the
radial coordinate using the discrete variable representation (DVR)
constructed from the positive energy Coulomb wave function. We show
that the Coulomb wave function DVR (CWDVR) is able to treat the
Coulomb singularity naturally and provide a good representation of
continuum wave functions.

The other advantage of CWDVR is that it needs 3-10 times fewer grid
points than FD scheme because of the uneven distribution of the grid
points, in which case one has a coarser grid at larger distances
where Coulomb potential plays a less important role and wave
functions oscillate less rapidly. Because the CWDVR is economic, it
is  a  promising step towards a more efficient treatment of the
many-electron systems which is extremely demanding~\cite{Meha05}. In
addition, many strong field processes, such as above threshold
ionization (ATI), high-order harmonic generation (HHG) and dynamical
stabilization etc.,  can be well understood within a single active
electron~(SAE) picture. Therefore, the hydrogenic atom serves a
prototype for spherically symmetrical atomic systems interacting
with intense laser fields.

The rest of the paper is organized as follows. In
section~\ref{sec1}, we give the details of  how we construct the
CWDVR, preceded by a brief  introduction to the general DVR method.
We then present in section~\ref{sec2} how we apply CWDVR to
discretize the TDSE for one-electron atomic systems in intense laser
fields. In section~\ref{sec3}, we  present some results for the
multiphoton ionization of H and H$^-$ as well as the short-time
dynamics of H in a static electric field. We show that ionization
rates calculated by the present economic and accurate method are in
excellent agreement with other theoretical calculations.  Finally,
we give a short conclusion.  Atomic units (a.u.) are used unless
otherwise specified.

\section{\label{sec1}Coulomb Wave Function DVR}

 The DVR method  has its origin in the transformation
method devised by Harris  and coworkers~\cite{Harris65} for
calculating matrix elements of complicated potential functions in a
truncated basis set. It was further developed by Dickinson and
Certain~\cite{Dickinson68} who showed the relationship between the
transformation method and the Gaussian quadrature rules for
orthogonal polynomials.  Light and coworkers~\cite{Lill82} first
explicitly used the DVR method as a basis representation for quantum
problems rather than just a means of evaluating Hamiltonian matrix
elements.   Ever since then, different types of  DVR methods have
found wide applications in different fields of physical and chemical
problems~\cite{Light00}. There continues to be many efforts to
construct new types of DVR and to apply DVR in the combination of
other numerical methods~\cite{Colb92}.

 Essentially, DVR is  a representation whose associated basis
 functions are localized about discrete values of the coordinate.
 DVR greatly simplifies the evaluation of Hamiltonian matrix elements.
 The potential matrix elements are merely the evaluation of interaction potential at the
 DVR  grid points and no integration is needed.
 The kinetic energy matrix elements can be calculated very simply and analytically
 in most cases~\cite{Baye86}.   In this section, we  first give
 a short introduction to the DVR constructed from the orthogonal
 polynomials. Then we present how we construct the DVR from the
 Coulomb wave function, which will be used to solve the TDSE of
 atomic systems in intense laser fields in section~\ref{sec2}.

\subsection{\label{sec1sub1}DVR Related to Orthogonal Polynomials}

It is well known~\cite{Baye86} that the DVR basis functions can be
constructed from any orthogonal polynomials,
$P_{N}(x)$,  defined in the domain $(a,b)$ with the corresponding weight function $%
\omega (x)$. Let
\begin{equation}
\mathcal{P}_{N}(x)=\sqrt{\omega (x) / h_{N}}P_{N}(x),
\label{calPx}
\end{equation}
where $h_{N}$ is the normalization constant such that
\begin{equation}
\int_{a}^{b}dx \mathcal{P}_{M}(x)\mathcal{P}_{N}(x)=\delta_{MN}.
\end{equation}
Then the cardinal function of $\mathcal{P}_{N}(x)$ is given
by~\cite{Boyd00}
\begin{equation}
\mathcal{C}_{i}(x)=\frac{1}{\mathcal{P}_{N}^{\prime }(x_{i})}\frac{\mathcal{P}_{N}(x)}{%
x-x_{i}},  \label{Cardinal1}
\end{equation}
in which  $x_{i}$ ($i=1,2,...,N$) are the zeros of $P_{N}(x)$, and
$\mathcal{P}_{N}^{\prime }(x_{i})$ stands for the first derivative
of $\mathcal{P}_{N}(x)$ at $x_{i}$. Apparently, $\mathcal{C}_{i}(x)$
satisfies the cardinality condition 
\begin{equation}
\mathcal{C}_{i}(x_j) = \delta_{ij}.
\label{Cardinality1}
\end{equation}

One can construct the DVR basis function $f_{i}(x)$ from the
cardinal function $\mathcal{C}_{i}(x)$ as follows:
\begin{equation}
f_{i}(x)=\frac{1}{\sqrt{\omega _{i}}}\mathcal{C}_{i}(x),
 \label{DVRbasis1}
\end{equation}
which, at the point $x_j$, gives
\begin{equation}
f_{i}(x_{j})=\frac{1}{\sqrt{\omega _{i}}}\delta _{ij}.
 \label{DVRbasiProp1}
\end{equation}

  We know that the integration of any function $F(x)$ can be calculated using
an appropriate quadrature rule  associated with the zeros of the
orthogonal polynomials, {\it i.e.}:
\begin{equation}
\int_{a}^{b}dxF(x)\simeq \sum_{i=1}^{N}\omega _{i}F(x_{i}),
 \label{GaussQuaRule1}
\end{equation}
where  $\omega _{i}$ is the corresponding weight at the point
$x_{i}$.
 From the theory of classical orthogonal polynomials \cite{Szego39}, the
integration formula Eq.~(\ref{GaussQuaRule1}) is exact as long as
the function $F(x)$ can be expressed as  a polynomial  of order
$2N-1$ (or lower) times the weight function $\omega(x)$. With the
help of Eqs.~(\ref{calPx}), (\ref{Cardinal1}) and~(\ref{DVRbasis1}),
it is easy to show that the function $f_{i}^{\ast}(x)f_{j}(x)$
satisfies this condition.  Therefore, the following integration can
be carried out exactly:
\begin{equation}
\int_{a}^{b}dxf_{i}^{\ast }(x)f_{j}(x)
=  \sum_{k=1}^{N}\omega_{k}f_{i}^{\ast}(x_{k})f_{j}(x_{k})
 =  \delta _{ij}.  \label{DVRbasiProp2}
\end{equation}

  As a result of Eqs.~(\ref{DVRbasiProp1}) and~(\ref{GaussQuaRule1}),
  any local operator $V(x)$ has a diagonal representation in the DVR
  basis set as follows:
\begin{equation}
\int_{a}^{b}dxf_{i}^{\ast }(x) V(x) f_{j}(x) \simeq
\sum_{k=1}^{N}\omega_{k}f_{i}^{\ast}(x_{k}) V(x_k)f_{j}(x_{k})
 =  V(x_i)\delta _{ij}.
 \label{Mtrxlocal}
\end{equation}
 On the contrary, the representation of  a differential
 operator in the DVR basis is usually full matrix. Nevertheless, in
 most cases, the matrix elements of the first and second differential operator
\begin{equation}
\int_{a}^{b}dxf_{i}^{\ast }(x)\frac{d}{dx}f_{j}(x),
\label{DVR1stDevr}
\end{equation}%
and
\begin{equation}
\int_{a}^{b}dxf_{i}^{\ast }(x)\frac{d^{2}}{dx^{2}}f_{j}(x),
\label{DVR2ndDevr}
\end{equation}
can  be  evaluated  analytically using the quadrature rule
Eqs.~(\ref{GaussQuaRule1}) or~(\ref{DVRbasiProp2}). Usually, the
final results are very simple expressions  of the zeros $x_{i}$ and
the number of the points $N$~\cite{Baye86}.

\subsection{\label{sec1sub2}DVR  Constructed from the Coulomb Wave Function}

Following a  similar spirit, we  present the  DVR basis function
$f_{i}(r)$ which is constructed from the Coulomb wave function. The
Coulomb wave function (see, Eq.~(14.1.1) of Ref.~\cite{Abra65})
satisfies the differential equation
\begin{equation}
\frac{d^{2}}{d\tilde{r}^{2}}v(\tilde{r})+\left[ 1-\frac{2\eta }{\tilde{r}}-%
\frac{L(L+1)}{\tilde{r}^{2}}\right] v(\tilde{r})=0,
\label{DiffCoulWav1}
\end{equation}
where  $\eta$ is real and $L$ is a non-negative integer.
Eq.~(\ref{DiffCoulWav1}) admits a regular solution $F_{L}\left( \eta
,\tilde{r}\right)$.

For our purpose, we will  consider the regular solution for $L=0$
and $\eta<0$. Denoting
\begin{equation}
\eta =-\frac{Z}{\sqrt{2E}},  \label{eta1}
\end{equation}
and
\begin{equation*}
\tilde{r}=r \sqrt{2E},
\end{equation*}
we can write Eq.~(\ref{DiffCoulWav1}) in another form as
\begin{equation}
\frac{d^{2}}{dr^{2}}v(r)=-\left[ 2E+\frac{2Z}{r}\right] v(r)\equiv W(r)v(r),
\label{DiffCoulWav2}
\end{equation}
whose solution  is given by
\begin{equation}
v(r)= F_{0}\left( -{Z}/{\sqrt{2E}}, r\sqrt{2E}\right) .
\label{solutvr}
\end{equation}

For any given energy $E$ and nuclear charge $Z$, the solution
(\ref{solutvr}) has simple zeros over $(0,\infty )$ (see Fig.~14.3
of Ref.~\cite{Abra65}). Similar to Eq.~(\ref{Cardinal1}), we can
define the cardinal function of $v(r)$ as
\begin{equation}
C_{i}(r)=\frac{1}{v^{\prime }(r_{i})}\frac{v(r)}{r-r_{i}}  \label{Cardinal2}
\end{equation}
where $r_{i}$ is the zero of $v(r)$, and $v^{\prime }(r_{i})$ stands
for its first derivative  at $r_{i}$.

Analogy to Eq.~(\ref{DVRbasis1}), one can construct the Coulomb wave
function DVR basis function
\begin{equation}
f_{i}(r)=\frac{1}{\sqrt{\omega _{i}}}C_{i}(r)=\frac{1}{\sqrt{\omega _{i}}}%
\frac{1}{v^{\prime }(r_{i})}\frac{v(r)}{r-r_{i}},  \label{DVRbasis2}
\end{equation}
which, at the zero $r_{j}$, equals to
\begin{equation}
f_{i}(r_{j})=\frac{1}{\sqrt{\omega _{i}}}\delta _{ij}.
\label{DVRbasiProp1pr}
\end{equation}

Following Schwartz \cite{Schw85}, Dunseath and coworkers
\cite{Duns02} constructed an appropriate  quadrature rule associated
with  the zeros $r_{i}$ of the Coulomb wave function
(\ref{solutvr}). By using this quadrature rule,  the  integration of
a function $F(r)$ over $(0,\infty )$ can be  evaluated as follows:
\begin{equation}
\int_{0}^{\infty }drF(r)\simeq \sum_{i=1}^{N}\omega _{i}F(r_{i}),
\label{GaussQuaRule2}
\end{equation}
where the weight $\omega _{i}$  is given by
\begin{equation}
\omega _{i}\simeq \frac{\pi }{a_{i}^{2}},  \label{weightR}
\end{equation}
with
\begin{equation}
a_{i} \equiv v^{\prime }(r_{i}).  \label{ai}
\end{equation}

Using the quadrature rule Eq.~(\ref{GaussQuaRule2}), we have the
orthogonality of the CWDVR basis functions
\begin{equation}
\int_{0}^{\infty }f_{i}^{\ast }(r)f_{j}(r)dr\simeq \sum_{k=1}^{N}\omega
_{k}f_{i}^{\ast }(r_{k})f_{j}(r_{k})=\delta _{ij}.
  \label{DVRbasiProp2pr}
\end{equation}
One can also evaluate the integration of the form (\ref{DVR1stDevr})
and (\ref{DVR2ndDevr}) using the same quadrature rule. We thus have
\begin{eqnarray}
P_{ij} &\equiv &\int_{0}^{\infty }f_{i}^{\ast }(r)\frac{d}{dr}f_{j}(r)dr
\label{Pij1} \\
&\simeq &\sum_{k=1}^{N}\omega _{k}f_{i}^{\ast }(r_{k})f_{j}^{\prime
}(r_{k}) =\sum_{k=1}^{N}\frac{\omega _{k}}{\sqrt{\omega _{i}\omega
_{j}}}\delta _{ik}C_{j}^{\prime }(r_{k}) \label{Pij1pr}
\end{eqnarray}
and
\begin{eqnarray}
T_{ij} &\equiv &-\frac{1}{2}\int_{0}^{\infty }f_{i}^{\ast }(r)\frac{d^{2}}{%
dr^{2}}f_{j}(r)dr  \label{Tij1} \\
&\simeq &-\frac{1}{2}\sum_{k=1}^{N}\omega _{k}f_{i}^{\ast
}(r_{k})f_{j}^{\prime \prime }(r_{k})
=-\frac{1}{2}\sum_{k=1}^{N}\frac{\omega _{k}}{\sqrt{\omega _{i}\omega _{j}}%
}\delta _{ik}C_{j}^{\prime \prime }(r_{k}). \label{Tij1pr}
\end{eqnarray}

It is easy to show  that~\cite{Schw85, Duns02}, for the solution
$v(r)$ to Eq.~(\ref{DiffCoulWav2}), the first and second derivative
of the cardinal function Eq.~(\ref{Cardinal2}) are respectively
given by
\begin{equation}
C_{j}^{\prime }(r_{k})=\left( 1-\delta _{jk}\right) \frac{a_{k}}{a_{j}}\frac{%
1}{r_{k}-r_{j}},  \label{Cpjk}
\end{equation}
 and
\begin{equation}
C_{j}^{\prime \prime }(r_{k})=\delta _{jk}\frac{c_{k}}{3a_{k}}-\left(
1-\delta _{jk}\right) \frac{a_{k}}{a_{j}}\frac{2}{\left( r_{k}-r_{j}\right)
^{2}},  \label{Cppjk}
\end{equation}
where $a_{k}$ is given by Eq.~(\ref{ai}) and $c_{k}$ is calculated
by
\begin{equation}
c_{k}=a_{k}W(r_{k}),  \label{ck}
\end{equation}
in which $W(r)=-\left[ 2E+\frac{2Z}{r}\right] $ (cf. Eq.~(\ref%
{DiffCoulWav2})).

Finally,  combining Eqs.~(\ref{Pij1})-(\ref{Cppjk}) gives us
\begin{equation}
P_{ij}=\left( 1-\delta _{ij}\right) \frac{1}{r_{i}-r_{j}},  \label{Pij2}
\end{equation}
and
\begin{equation}
T_{ij}=-\delta _{ij}\frac{c_{i}}{6a_{i}}+\left( 1-\delta _{ij}\right) \frac{1%
}{\left( r_{i}-r_{j}\right) ^{2}}.  \label{Tij2}
\end{equation}
 with the help of Eq.~(\ref{weightR}).

Note that Eq.~(\ref{DVRbasiProp2pr}) is satisfied approximately for
the DVR basis
 functions constructed from the Coulomb wave function, while
 Eq.~(\ref{DVRbasiProp2}) is satisfied  exactly for the DVR basis  functions
 constructed from the
 orthogonal polynomial.  Nevertheless, we  will show later that CWDVR can
 be applied to solve the time-dependent Schr\"odinger equation very
 efficiently and accurately.

\section{\label{sec2}One-electron atomic system  in intense laser fields}

Let us consider an effectively one-electron atomic system in a laser
field. In spherical coordinates, the time-dependent Schr\"{o}dinger
equation is given by
\begin{equation}
i\frac{\partial }{\partial t}\Psi \left( \mathbf{r},t\right) =\left[
H_{0}(\mathbf{r})+H_{I}(\mathbf{r},t)\right] \Psi \left(
\mathbf{r},t\right) ,  \label{tdse0}
\end{equation}
in which the field free Hamiltonian $H_{0}$ is defined by
\begin{eqnarray}
H_{0} &=&-\frac{1}{2}\nabla ^{2}+V_{C}^l(r)  \notag \\
&=&-\frac{1}{2}\left[ \frac{1}{r^{2}}\frac{\partial }{\partial r}\left( r^{2}%
\frac{\partial }{\partial r}\right)
-\frac{1}{r^{2}}\hat{L}^{2}\right] +V_{C}^l(r),  \label{Hfree}
\end{eqnarray}
where $V_{C}^l(r)$ is the effective Coulomb potential or any kind of
short (or zero) range model potential, which can  depend on the
angular momentum number $l$.  And the orbital angular momentum
operator $\hat{L}^{2}$ is defined by
\begin{equation}
\hat{L}^{2}=-\frac{1}{\sin \theta }\frac{\partial }{\partial \theta }\left(
\sin \theta \frac{\partial }{\partial \theta }\right) -\frac{1}{\sin
^{2}\theta }\frac{\partial ^{2}}{\partial \phi ^{2}},
\label{Lsqr}
\end{equation}
which satisfies the eigenvalue equation
\begin{equation}
\hat{L}^{2}Y_{lm}\left( \theta ,\phi \right) =l\left( l+1\right)
Y_{lm}\left( \theta ,\phi \right) ,  \label{Lsqr1}
\end{equation}
with $Y_{lm}\left( \theta ,\phi \right) $ being the spherical
harmonics.

In Eq.~(\ref{tdse0}), the interaction Hamiltonian $H_{I}(\mathbf{r}%
,t) $ describes the interaction of the active electron with the
applied laser pulse. In the laser parameters range of interest in
the present work, the dipole approximation  is well justified. For a
linearly polarized laser along the $z$ axis, $H_{I}(\mathbf{r},t)$
is given by
\begin{equation}
H_{I}^{(L)}(\mathbf{r},t)=\mathbf{r}\cdot \mathbf{E}(t)=r\cos \theta
E(t), \label{interLen1}
\end{equation}
in the length gauge, and
\begin{equation}
H_{I}^{(V)}(\mathbf{r},t)=-i\frac{1}{c}\mathbf{A}(t)\cdot
\mathbf{\nabla} =-i\frac{1}{c}A(t)\nabla _{0},  \label{interVel1}
\end{equation}
 the velocity gauge, where $c$ is the speed of light in the vacuum. $\nabla
_{0}$ denotes the $0th$ spherical component of the gradient operator
${\nabla}$~\cite{Vars88}. The
electric field strength $\mathbf{E}(t)$ is related to the vector potential $%
\mathbf{A}(t)$ of the laser pulse by
\begin{equation}
\mathbf{E}(t)=-\frac{1}{c}\frac{\partial }{\partial t}\mathbf{A}(t).
 \label{Et2At}
\end{equation}

\subsection{\label{sec2sub1}Discretization of  the Spatial Coordinates}
  In order solve the TDSE~(\ref{tdse0}), we need to discretize this
  equation. We
  expand the angular part of the wave function in terms of spherical harmonics. The
  radial coordinate $r$ can be discretized in different ways. The
  most straightforward one is to use the finite difference  method, in
  which case the first and second derivative with respect to $r$
  in Eq.~(\ref{Hfree}) are approximated by   formulas involving
  only several neighboring points. In the present work, we expand
  the radial part of the wave function in CWDVR basis functions constructed above.
  As seen from Eqs.(~\ref{Pij2}) and~(~\ref{Tij2}), CWDVR is a
  global method, in which case the representations of the derivatives
  involve all the grid points.

\subsubsection{Expansion of  the Angular Part}

Expanding the time dependent wave function in the following form
\begin{equation}
\Psi \left( \mathbf{r},t\right)\equiv \Psi \left( r,\theta, \phi,
t\right)
=\sum_{l=0}^{L}\sum_{m=-L}^{L}\frac{%
\varphi _{lm}(r,t)}{r}Y_{lm}\left( \theta ,\phi \right),
\label{expan1}
\end{equation}
and substituting into Eq.~(\ref{tdse0}), we arrive at
\begin{eqnarray}
&&\sum_{l=0}^{L}\sum_{m=-L}^{L}\frac{1}{r}Y_{lm}\left( \theta ,\phi \right) i%
\frac{\partial }{\partial t}\varphi _{lm}(r,t)  \notag \\
&=&\sum_{l=0}^{L}\sum_{m=-L}^{L}\frac{1}{r}Y_{lm}\left( \theta ,\phi \right) %
\left[ -\frac{1}{2}\frac{\partial ^{2}}{\partial r^{2}}+\frac{l\left(
l+1\right) }{2r^{2}}+V_{c}^l(r)\right] \varphi _{lm}(r,t)  \notag \\
&&+\sum_{l=0}^{L}\sum_{m=-L}^{L}H_{I}(\mathbf{r},t)\frac{1}{r}%
\varphi _{lm}(r,t)Y_{lm}\left( \theta ,\phi \right),
  \label{tdse1}
\end{eqnarray}
with the help of Eq.~\ref{Lsqr1}.

 Multiplying the above equation by
$Y_{l^{\prime}m^{\prime}}^*\left( \theta ,\phi \right) /{r} $ and
integrating over $r^{2}\sin \theta d\theta d\phi $ on both sides, we
can rewrite it  as
\begin{eqnarray}
i\frac{\partial }{\partial t}\varphi _{l^{\prime }m^{\prime }}(r,t) &=&-%
\frac{1}{2}\frac{d^{2}}{dr^{2}}\varphi _{l^{\prime }m^{\prime
}}(r,t)+V_{eff}^l\left( r\right) \varphi _{l^{\prime }m^{\prime
}}(r,t)  \notag
\\
&&+\left[ H_{I}(r,t)\right] _{l^{\prime }m^{\prime }}^{\varphi},
\label{tdse2}
\end{eqnarray}
where we have made use of
\begin{equation}
\int_{0}^{\pi }\sin \theta d\theta \int_{0}^{2\pi }d\phi
Y_{l^{\prime }m^{\prime }}^{\ast }\left( \theta ,\phi \right)
Y_{lm}\left( \theta ,\phi \right) =\delta _{l^{\prime }l}\delta
_{m^{\prime }m}.  \label{OrthSpher}
\end{equation}

In Eq.~(\ref{tdse2}), we have defined an effective potential term
\begin{equation}
V_{eff}^l(r)\equiv V_{C}^l(r)+\frac{l(l+1)}{2r^{2}},
\label{EffPoten}
\end{equation}
and  a laser-interaction-related term,
\begin{eqnarray}
\left[ H_{I}(r,t)\right] _{l^{\prime }m^{\prime }}^{\varphi }
&\equiv&\sum_{l=0}^{L}\sum_{m=-L}^{L}\int_{0}^{\infty
}r^{2}dr\int_{0}^{\pi }\sin \theta d\theta \int_{0}^{2\pi }d\phi
\notag \\ &\times& \frac{1}{r}Y^*_{l^{\prime }m^{\prime }}\left(
\theta ,\phi \right) H_{I}(\mathbf{r},t)\frac{1}{r}\varphi
_{lm}(r,t)Y_{lm}\left( \theta ,\phi \right). \label{Hinter1}
\end{eqnarray}

In the length gauge, substituting Eq.~(\ref{interLen1}) into Eq.~(\ref{Hinter1}%
) and making use of Eq.~(\ref{OrthSpher}) and the following
formula~\cite{Vars88}
\begin{equation}
\cos \theta Y_{lm}\left( \theta ,\phi \right) =a_{l+1m}Y_{l+1m}\left( \theta
,\phi \right) +a_{lm}Y_{l-1m}\left( \theta ,\phi \right) ,  \label{cosY}
\end{equation}
where
\begin{equation}
a_{lm}=\sqrt{\frac{\left( l-m\right) \left( l+m\right) }{\left( 2l-1\right)
\left( 2l+1\right) }},  \label{CoeAlm}
\end{equation}
we arrive at
\begin{equation}
\left[ H_{I}^{(L)}(r,t)\right] _{l^{\prime }m^{\prime }}^{\varphi }=r E(t)%
\left[ a_{l^{\prime }m^{\prime }}\varphi _{l^{\prime }-1m^{\prime
}}(r,t)+a_{l^{\prime }+1m^{\prime }}\varphi _{l^{\prime }+1m^{\prime }}(r,t)%
\right] .  \label{HPsiLen1}
\end{equation}

In the velocity gauge, substituting Eq.~(\ref{interVel1})  into
Eq.~(\ref{Hinter1}) and making use of Eq.~(\ref{OrthSpher}) and the
formula~\cite{Vars88}
\begin{eqnarray}
\nabla _{0}\left[ \frac{1}{r}R(r)Y_{lm}\left( \theta ,\phi \right) \right]
&=&a_{l+1m}Y_{l+1m}\left( \theta ,\phi \right) \frac{1}{r}\left( \frac{d}{dr}%
-\frac{l+1}{r}\right) R(r)  \notag \\
&&+a_{lm}Y_{l-1m}\left( \theta ,\phi \right) \frac{1}{r}\left( \frac{d}{dr}+%
\frac{l}{r}\right) R(r),  \label{Delt0RY}
\end{eqnarray}
 we arrive at
\begin{eqnarray}
\left[ H_{I}^{(V)}(r,t)\right] _{l^{\prime }m^{\prime }}^{\varphi } &=&iA(t)%
\frac{1}{r}\left[ l^{\prime }a_{l^{\prime }m^{\prime }}\varphi _{l^{\prime
}-1m^{\prime }}(r,t)-\left( l^{\prime }+1\right) a_{l^{\prime }+1m^{\prime
}}\varphi _{l^{\prime }+1m^{\prime }}(r,t)\right]  \notag \\
&&-iA(t)\frac{d}{dr}\left[ a_{l^{\prime }m^{\prime }}\varphi
_{l^{\prime }-1m^{\prime }}(r,t)+a_{l^{\prime }+1m^{\prime }}\varphi
_{l^{\prime }+1m^{\prime }}(r,t)\right],  \label{HPsiVel1}
\end{eqnarray}
with $a_{lm}$ given by Eq.~(\ref{CoeAlm}).

\subsubsection{Discretization of the Radial Coordinate}

We have now changed the TDSE~(\ref{tdse0}) into~(\ref{tdse2}) with
the interaction term $\left[ H_{I}(r,t)\right]_{l^{\prime }m^{\prime
}}^{\varphi}$ given by Eq.~(\ref{HPsiLen1}) in the length gauge and by Eq.~(\ref{HPsiVel1}%
) in the velocity gauge.     Although we deal with the $r$
coordinate using CWDVR in the present work, we still give  the FD
formulas  for the purpose of comparison.  In the FD scheme,   the
grid points are chosen to be equal spacing $\Delta r$ such as
\begin{equation}
r_{i}=i\Delta r\text{, \ \ }i=1,2,...,N.  \label{FDRpts}
\end{equation}

The 5-point central finite difference approximation to the first and
second derivative of $\varphi _{lm}(r,t)$ are given by
\begin{eqnarray}
\frac{d}{dr}\varphi _{lm}(r,t)&=&\frac{1}{12\Delta r}\left[ \varphi
_{lm}(r-2\Delta r,t)-8\varphi _{lm}(r-\Delta r,t) \right. \nonumber
\\
&& \left. +8\varphi _{lm}(r+\Delta r,t)-\varphi _{lm}(r+2\Delta
r,t)\right], \label{FD1stDev}
\end{eqnarray}
and
\begin{eqnarray}
\frac{d^{2}}{dr^{2}}\varphi _{lm}(r,t) &=&-\frac{1}{12\left( \Delta r\right)
^{2}} \left[ \varphi _{lm}(r-2\Delta r,t)-16\varphi _{lm}(r-\Delta r,t) \right.
\nonumber \\
&&\left.+30\varphi _{lm}(r,t)-16\varphi _{lm}(r+\Delta r,t)+\varphi
_{lm}(r+2\Delta r,t)\right].  \label{FD2ndDev}
\end{eqnarray}

In order to have give a better ground state energy, we have used the
following approximation for the second derivative at the first point
to take into account the boundary condition:
\begin{eqnarray}
\frac{d^{2}}{dr^{2}}\varphi _{lm}(r,t) &=&  -\frac{1}{12\left(\Delta
r\right)^{2}} \left[ 30\varphi _{lm}(r,t)-16\varphi _{lm}(r+\Delta
r,t)+\varphi _{lm}(r+2\Delta r,t)\right]
\nonumber \\
&& + C_0 \varphi _{lm}(r,t), \label{FD2ndDevPr}
\end{eqnarray}
where $C_0$ is a constant which depends on the grid spacing $\Delta
r$.  For example, we take $C_0=-1.48986$  for $\Delta r = 0.2$ and
get a converged  H ground state energy of  $-0.500000065$ a.u., but
$C_0=-1.814116$  for $\Delta r = 0.3$  in order to get the same
value of the ground state energy.

  Apparently, one needs a very small spacing $\Delta r$ in order to
have a good approximation to these derivatives. For the atomic
systems interacting with the intense laser pulses, the electronic
wave function can  be generally driven to hundreds of or even
thousands of atomic unit away from the nuclear. Therefore, we need a
very large number of the grid points. Another disadvantage of the FD
scheme is that, one needs to carefully deal with the Coulomb
singularity at the origin~\cite{Bauer05}.

Now let us turn to   discretize $r$ using the CWDVR basis functions
which we discussed in the last  section. As we mentioned before, the
CWDVR has several advantages: firstly, it deals with the singularity
of the Coulomb-type potential at the $r=0$ naturally; secondly, the
grid points ({\it i.e.}, zeros of Coulomb wave function) are dense
nearly the origin where the Coulomb potential plays  a crucial role
and sparse at large distances where it is not very important;
thirdly, compared to FD scheme, much fewer grid points are needed
for the same extent of the grid because of the uneven distribution
of the grid points. 

The CWDVR basis functions $f_{i}(r)$  are given in
Eqs.~(\ref{DVRbasis2}) , (\ref{DVRbasiProp1pr}) and
(\ref{DVRbasiProp2pr}). Let us start from Eq.~(\ref{tdse2}) and
expand $\varphi _{lm}(r,t)$ in terms of \ $f_{i}(r)$ as follows:
\begin{equation}
\varphi _{lm}(r,t)=\sum_{i=1}^{N}D_{ilm}(t)f_{i}(r),  \label{ExpanPhilm}
\end{equation}
where the coefficient is given by,
\begin{eqnarray}
D_{ilm}(t) &=&\int_{0}^{\infty }drf^*_{i}(r)\varphi _{lm}(r,t)  \notag \\
&\simeq &\sum_{k=1}^{N}\omega _{k}f^*_{i}(r_{k})\varphi
_{lm}(r_{k},t)  \notag
\\
&=&\sqrt{\omega _{i}}\varphi _{lm}(r_{i},t).  \label{CoefDilm}
\end{eqnarray}

Substituting Eq.~(\ref{ExpanPhilm}) into Eq.~(\ref{tdse2}),
multiplying both sides by $f_{i^{\prime }}^{\ast }(r)$ and
integrating over $r$, we arrive at
\begin{eqnarray}
i\frac{\partial }{\partial t}D_{i^{\prime }l^{\prime }m^{\prime }}(t)
&=&\sum_{i=1}^{N}T_{i^{\prime }i}D_{il^{\prime }m^{\prime
}}(t)+V_{eff}\left( r_{i^{\prime }}\right) D_{i^{\prime }l^{\prime
}m^{\prime }}(t)  \notag \\
&&+\left[ H_{I}(r,t)\right] _{i^{\prime }l^{\prime }m^{\prime }}^{\varphi }
\label{tdse3}
\end{eqnarray}
where we have made use of Eq.~(\ref{Tij1}).   $\left[
H_{I}(t)\right] _{i^{\prime }l^{\prime }m^{\prime }}^{\varphi }$
stands for the matrix element of the interaction term, which  is
shown to be
\begin{eqnarray}
\left[ H_{I}^{(L)}(t)\right] _{i^{\prime }l^{\prime }m^{\prime
}}^{\varphi } &=&\sum_{i=1}^{N}\int_{0}^{\infty }drf_{i^{\prime
}}(r)f_{i}(r)E(t)r\left[ a_{l^{\prime }m^{\prime }}D_{il^{\prime
}-1m^{\prime }}(t)+a_{l^{\prime
}+1m^{\prime }}D_{il^{\prime }+1m^{\prime }}(t)\right]   \notag \\
&=&E(t)r_{i^{\prime }}\left[ a_{l^{\prime }m^{\prime }}D_{il^{\prime
}-1m^{\prime }}(t)+a_{l^{\prime }+1m^{\prime }}D_{il^{\prime }+1m^{\prime
}}(t)\right] .  \label{HPsiLen2}
\end{eqnarray}
in the length gauge. In the velocity gauge, it takes a slightly
complicated form as
\begin{eqnarray}
\left[ H_{I}^{(V)}(t)\right] _{i^{\prime }l^{\prime }m^{\prime
}}^{\varphi
} &=&iA(t)\sum_{i=1}^{N}\int_{0}^{\infty }drf_{i^{\prime }}(r)\frac{1}{r}%
f_{i}(r)\left[ l^{\prime }a_{l^{\prime }m^{\prime }}D_{il^{\prime
}-1m^{\prime }}(t)-\left( l^{\prime }+1\right) a_{l^{\prime }+1m^{\prime
}}D_{il^{\prime }+1m^{\prime }}(t)\right] \notag \\
&&-iA(t)\sum_{i=1}^{N}\int_{0}^{\infty }drf_{i^{\prime }}(r)\frac{d}{dr}%
f_{i}(r)\left[ a_{l^{\prime }m^{\prime }}D_{il^{\prime }-1m^{\prime
}}(t)+a_{l^{\prime }+1m^{\prime }}D_{il^{\prime }+1m^{\prime }}(t)\right] \notag \\
&=&iA(t)\frac{1}{r_{i^{\prime }}}\left[ l^{\prime }a_{l^{\prime
}m^{\prime }}D_{i^{\prime }l^{\prime }-1m^{\prime }}(t)-\left(
l^{\prime }+1\right) a_{l^{\prime }+1m^{\prime }}D_{i^{\prime
}l^{\prime }+1m^{\prime }}(t)\right] \notag
\\
&&-iA(t)\sum_{i=1}^{N}P_{i^{\prime }i}\left[ a_{l^{\prime }m^{\prime
}}D_{il^{\prime }-1m^{\prime }}(t)+a_{l^{\prime }+1m^{\prime
}}D_{il^{\prime }+1m^{\prime }}(t)\right].
 \label{HPsiVel2}
\end{eqnarray}
 where we have made use of Eq.~(\ref{Pij1}).  The matrix elements
  of   $P_{i^{\prime }i}$ in
 Eq.~(\ref{HPsiVel2}) and  $T_{i^{\prime }i}$ in Eq.~(\ref{tdse3})
  are calculated analytically from
 Eqs.~(\ref{Pij2}) and~(\ref{Tij2}) respectively.
 The zeros $r_i$ needed for evaluating these matrix elements are calculated with
 the help of COULFG~\cite{Barn82}. Please note
 that, for a linearly polarized laser considered in the present
 work, the subscript index $m^{\prime}$ is equal to $0$. In this
 case, we only have  two dimensional matrices with indexes $i^{\prime}$ and
 $l^{\prime}$.

\subsubsection{Distribution of the CWDVR Grid Points}
 \begin{table}[t]
\caption{Comparison of the grid points distribution of  CWDVR for
 different $Z$  and  $\kappa$.}
\begin{ruledtabular}
\begin{tabular}{ll|cccccc}
& &$\kappa=0.5$&$\kappa=1.0$&$\kappa=2.0$&$\kappa=3.0$&$\kappa=4.0$
&$\kappa=5.0$\\
\hline
 $Z$=12 &$N$&47&64&106&152&198&245 \\
  &$r_1$&0.1529&0.1526&0.1517&0.1501&0.1480&0.1455 \\
  &$r_N$&153.29&150.70&150.04&150.86&150.45&150.52 \\
 &$\Delta r_0$&0.3589&0.3565&0.3472&0.3335&0.3171&0.2994 \\
 &$\Delta r_{15}$&2.3650&1.9090&1.3227&0.9655&0.7491&0.6091 \\
 &$\Delta r_{150}$&4.9116&2.9158&1.5402&1.0381&0.7815&0.6263 \\
\hline
   $Z$=20 &$N$&56&72&112&156&202&248 \\
  &$r_1$&0.09174&0.09169&0.09148&0.09114&0.09066&0.09007 \\
  &$r_N$&150.27&151.39&150.74&150.47&150.73&150.43 \\
 &$\Delta r_0$&0.2157&0.2151&0.2130&0.2097&0.2053&0.2001 \\
 &$\Delta r_{15}$&1.8308&1.6554&1.2137&0.9132&0.7270&0.5972\\
 &$\Delta r_{150}$&4.3561&2.7914&1.5209&1.0320&0.7789&0.6250
\end{tabular}
\end{ruledtabular}
\label{table1}
\end{table}
The grid point $r_i$ of CWDVR  is the solution  of $v(r)=0$ where
$v(r)$ is defined by Eq.~(\ref{solutvr}) for  any given energy $E$
and nuclear charge $Z$. Therefore, the distribution of the  CWDVR
grid points
  can be adjusted by the
value of parameter $Z$ and $\kappa\equiv\sqrt{2E}$.

In table~\ref{table1}, we compare  the grid points distribution of
the CWDVR for the same maximum grid point value $r_{max}\simeq150$
and for different $\kappa$ values ranging from 0.5 to 5. The
parameter  $Z$ takes  either  12 or 20.  We list in this table the
number of the grid points $N$ (up to the first point which is
greater than 150) and  the value of the first and the last grid
point $r_1$ and $r_N$. We also give three grid spacings $\Delta
r_0$, $\Delta r_{15}$ and $\Delta r_{150}$ which correspond to the
spacing between the first two points, the two points around 15 and
the last two grid points.

One notices that the value of $Z$  mainly determines the value of
the first grid point $r_1$, {i.e.}, the greater $Z$ is, the smaller
$r_1$ becomes. However, increasing the value of $\kappa$ will mainly
decrease the spacing between the grid points at large $r$ and thus
one needs more grid points for the same value of $r_{max}=150$. At
the same time, larger $\kappa$ means much more even distribution at
larger distances. For example, $\Delta r_{15}$  and $\Delta r_{150}$
for $\kappa=5$ differ much less than that for $\kappa=0.5$.

For the same $r_{max}=150$ a.u., one observes that  the number of
the grid points for CWDVR cases is only 1/10 to 1/3 of that for FD
case if we choose even spacing $\Delta r = 0.2$ a.u.. As we will
discuss later, most of our results for the hydrogen are convergent
for $\kappa=1.0$  and  $Z=20$ case where only 1/10 of the number of
the grid points for FD is needed ($N=72$ vs $N=750$).

\subsection{\label{sec2sub2}Wave Function Propagation in Time}

After the  discretization of the spatial coordinates, one has to
advance the evolution of  the initial wave function in discretized
time. If the initial wave function one desires is  the electronic
ground state, it can be computed by the  propagation of any trial
wave function in imaginary time. The Schr\"odinger equation  in
imaginary time is mapped into the diffusion equation. In this case,
any excited states will decay away faster than the ground state
since the latter has the lowest energy. Once the energy is adjusted
to the true ground state energy $\varepsilon_0$, the asymptotic
solution is a steady-state solution. One will thus obtain a
converged ground state energy and wave function on the spatial grids
after a sufficiently long time of diffusion.

Concerning the propagation of the time-dependent Schr\"odinger
equation, there exist many different methods such as split
operator~\cite{Feit83}, Chebychev polynomial expansion~\cite{Tal84},
Taylor series~\cite{Qiao02} and Arnoldi/Lanczos
method~\cite{Park86}.  There have been a number of authors who made
a detailed comparison study of the efficiency and accuracy among
different propagation schemes~\cite{Lefo91}.  Although the choice of
the propagation scheme  depends on the internal characteristics of
the physical problem at hand, it is generally accepted that
Arnoldi/Lanczos method proves to offer an accurate and flexible
approximation of the matrix exponentials involved in the propagation
of wave function for most practical applications (Note that Arnoldi
method can apply to nonsymmetric matrices, but Lanczos method only
applies to symmetric or Hermitian matrices).

In the  Arnoldi/Lanczos method, the eigenvalue and eigenstate of a
large matrix is approximated in the Krylov subspace spanned by the
vectors $\{\psi_0, H\psi_0, H^2\psi_0,...,H^m\psi_0\}$ for the order
$m$ which is typically much smaller than the dimension of the
Hamiltonian matrix $H$ itself. In practice, we need to use
Arnoldi/Lanczos algorithm~\cite{Saad03} to generate $m$ orthonormal
basis vectors $V_{m+1}$ in the Krylov subspace. At the same time, an
upper Hessenberg matrix $h_{m+1}$ is formed resulting from these
processes. The eigenvalues of the original large matrix are thus
approximated by the eigenvalues of a Hessenberg matrix of dimension
$m+1$.  Therefore, Arnoldi/Lanczos method is especially suitable for
the large matrix and multi-dimensional problems.

 An extensive software package ``Expokit''  for
the computation of matrix exponential was developed by
Sidje~\cite{Sidj98}.  In this package, the author has provided
several alternatives to compute the matrix-valued function $e^{tH}$
for small, dense complex matrices $H$. In addition, ``Expokit'' has
functions for computing $e^{tH}\psi_0$ for both small, dense
matrices and large, sparse matrices.

In the present work, we have incorporated  this software into our
code for the accurate propagation of the wave function for atomic
systems interacting with strong laser pulses.  In each  time step,
we first generate the orthonormal vectors $V_{m+1}$ and the upper
Hessenberg matrix $h_{m+1}$ for the wave function $\Psi(t)$ using
the Arnoldi process. Then the exponential of the upper Hessenberg
matrix  $ e^{-i\Delta th_{m+1}} $   is carried out by the
irreducible rational Pad\'e approximation combined with
scaling-and-squaring. Finally, the wave function at the next time
step is given by
 \begin{equation}
  \Psi(t+\Delta t)  = |\Psi(t)| V_{m+1} e^{-i\Delta t h_{m+1}}e_1,
  \label{WvProp}
 \end{equation}
where   $|\Psi(t)|$ stands for the norm of the wave function at time
$t$ and $e_1$ is the first unit basis vector~\cite{Sidj98}.

For all the results presented in this paper, we use the Arnoldi
order $m=30$ and the propagation time step $\Delta t=0.01$ a.u..
However, we find that for some of  the laser parameters we consider,
the results are already converged for much lower order of $m$ (e.g.,
12) and much larger time step $\Delta t$ (e.g., 0.04 a.u.).

\section{\label{sec3}Results and Discussions}

In this section, we will present some numerical results for
ionization of the hydrogen atom by  strong static electric fields
and by intense laser fields. In order to illustrate that our method
can equally apply to any effectively one-electron atomic system with
a proper SAE model potential, we have also calculated multiphoton
detachment rates of the negative hydrogen ion for several laser
frequencies at different intensities.  Excellent agreements are
achieved between our results and the previous theoretical
calculations.

  For an  accurate evaluation of the ionization rate for a particular
laser intensity, the laser pulse should ramp on adiabatically   and
keep constant for a sufficiently long time.  The laser pulse
ramping-on time must be large compared to the initial atomic orbital
period of the bound electron and thus the electron energy will
adjust adiabatically to the rising intensity. The constant-intensity
time should also be sufficiently long so that the frequency
bandwidth is small compared with the laser frequency. Only in this
case, we can treat the laser field as monochromatic. Taking these
into consideration, we take the vector potential $\mathbf{A}(t)$ of
the laser pulse having the following form:
\begin{equation}
   \mathbf{A}(t)= A_0 f(t) \cos (\omega t)\hat{\mathbf{k}},
\end{equation}
where the pulse envelope, $f(t)$, is given by 
\begin{equation}
   f(t)= \left\{
   \begin{array}{lc}
      \frac{1}{2}\left[1-\cos\left(\frac{\pi t}{\tau_1}\right)\right], &
      0\leq t\leq \tau_1,\\
      1, & \tau_1\leq t \leq \tau_1+\tau_2,
  \end{array}
  \right.
  \label{AtEnvlop}
\end{equation}
where $\tau_1$ and $\tau_2$ are normally taken to be 5-10 and 10-40
 laser cycles, depending on the laser frequency under
consideration. The peak value of the vector potential $A_0$ is
connected to the peak laser intensity $I_0$ by
  \begin{equation}
 A_0= \frac{E_0} {\omega}= \frac{1}{\omega} \sqrt{\frac{I_0}{I_{au}}},
 \label{PeakA0}
\end{equation}
where  the atomic unit of laser intensity $I_{au}$ equals to
$3.5094\times10^{16}$ W/cm$^2$. The electric field strength
 $\mathbf{E}(t)$ is calculated from Eq.~(\ref{Et2At}).

 In order to avoid the reflection of wave function by the edge of
 the $r$ grid,  the wave function is  multiplied  by an absorbing
 function in each step of the propagation as follows
  \begin{equation}
   \Psi(\mathbf{r}, t)  = M(r) \Psi(\mathbf{r}, t),
   \label{WvSplit}
   \end{equation}
 in which
 \begin{equation}
   M(r)= \left\{
   \begin{array}{lc}
      1, &
      r\leq r_{\alpha},\\
      exp\left[ -\left(\frac{r - r_{\alpha}}{r_{\sigma}}\right)^2\right],
      & r>r_{\alpha},
  \end{array}
  \right.
  \label{MaskFun}
\end{equation}
 where $r_{\alpha} = {\alpha} r_{max}$ and  $r_{\sigma} = {\sigma} r_{max}$ with $r_{max}$ being
 the maximum value of the grid. It is very important to carefully choose
 the absorbing parameters $\alpha$ and $\sigma$ such that  the function $M(r)$
 is sufficiently smooth and that the wave function near the edge
 is completely absorbed without any
 reflection.  It is extremely crucial
 to avoid any unphysical effects induced by the absorbing potential~\cite{Riss95}.
  According to our experiences,   we take $0.3\le\alpha\le0.6$ and
 $0.5\le\sigma\le5.0$ depending on the value of $r_{max}$ and the laser
 parameters. The  criteria is that the time-dependent physical
 quantities,
 such as the population decay of the ground state, should be  converged
 against any small variation of these parameters in the vicinities of some
 values.

  Of course, convergence
has to be achieved against changing of other parameters such as the
largest quantum number of the angular momentum $L$, the number of
grid points  $N$ in $r$ coordinate, the propagation time step
$\Delta t$ and the Arnoldi order $m$ etc..

\subsection{\label{sec3sub1}Choice of Gauge and   Convergence of the CWDVR Grid }
\begin{figure}
    \includegraphics[width=8cm]{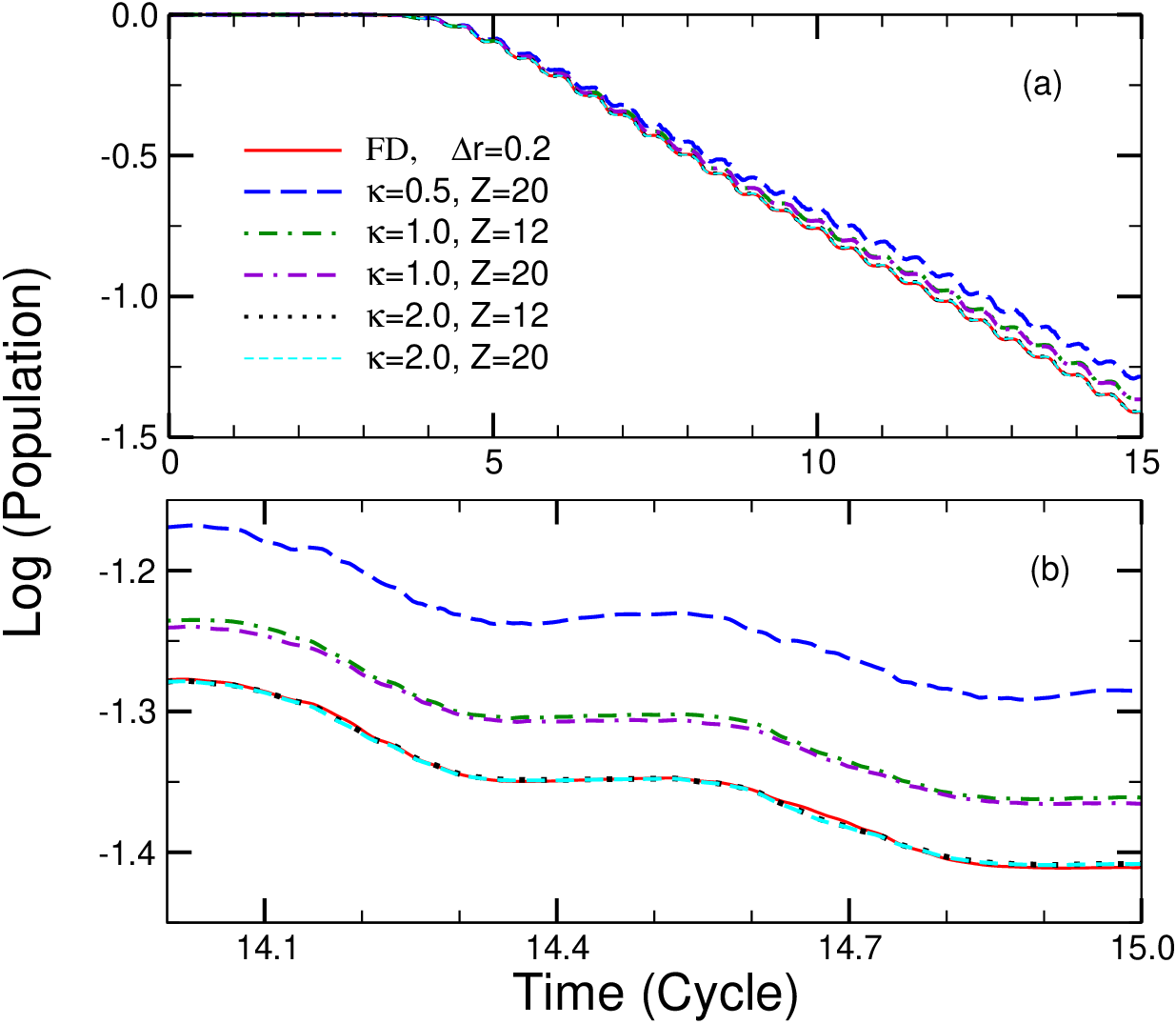}\\
    \caption{(color online). Convergence of the CWDVR grid for different
    $\kappa$ and $Z$ values. The natural logarithm of the
    population  decay as a function of time (in unit of laser cycle)
    are shown for the laser wavelength 780 nm at the peak intensity of
    $2\times10^{14}$ W/cm$^{2}$.   (a) Results from different CWDVR
    grids ($\kappa$ and $Z$ as indicated) are compared against that from
    FD method; (b) A magnified  version of (a) for the time from 14 to 15
    cycle.}
    \label{figure1}
\end{figure}
 Although different gauges describing the
 interacting Hamiltonian $H_I(t)$ should  in principle lead to the same physical
 results, it is not true in practice because of different approximations
 and inaccurate numerical wave functions. For some approximate methods,
 length gauge proves to be better than the velocity gauge
 in some circumstances~\cite{Kjel05}.
 However,
 as discussed by  Cormier and Lambropoulos~\cite{Corm96}, velocity gauge is
 preferable for some
  {\it ab initio} numerical calculations. This is connected with the
 fact that the canonical momentum in the velocity gauge has been
 reduced to a slowly varying variable since the momentum due to
 the strong field has been mainly removed~\cite{Corm96}. In this
 case, we can  avoid some widely varying variables. Especially in
 the present case where the wave function is expanded in terms of
 the spherical harmonics, much smaller $L$ is needed for a converged result
  in the velocity
 gauge than that needed in the length gauge.

 In the present work, we have performed our results in
 the velocity gauge for H in laser fields. However,
 length gauge is used for H in static
 field and H$^-$ in laser fields. One of the reasons
 for the latter choice  is that it is easy for us to compare our results with
 the previous results. Another reason, for $H^-$, is that the
 angular-momentum-dependent model potential~\cite{Laug93} makes the form of the
 potential in the velocity gauge  is not able to be obtained obviously~\cite{Teln95}.

Since the CWDVR grid depends on the two parameters $Z$ and $\kappa$,
one has to make sure that any calculation should be converged
against changes of them.  As shown in table~\ref{table1}, the CWDVR
grid for smaller $\kappa$ is much coarser.  At the same time, the
first point of the grid is mainly decided by the value of $Z$. It is
expected that much finer CWDVR grid will have a better
representation of the Coulomb potential and the wildly driven
electronic wave function by the laser field.

In order to test the convergence of CWDVR grid, the multiphoton
ionization of H for an infrared laser at a  moderate laser intensity
serves as a good example. For a hydrogenic atom, the potential is
given by
 \begin{equation}
V_C(r) = -\frac{Z}{r}
 \end{equation}
where $Z$ is the nuclear charge and equals to 1 for H.  For the
CWDVR grids we consider in table~\ref{table1}, we get the H  ground
state energy of  $-0.49999997$ a.u. or better.
\begin{figure}
    \includegraphics[width=8cm]{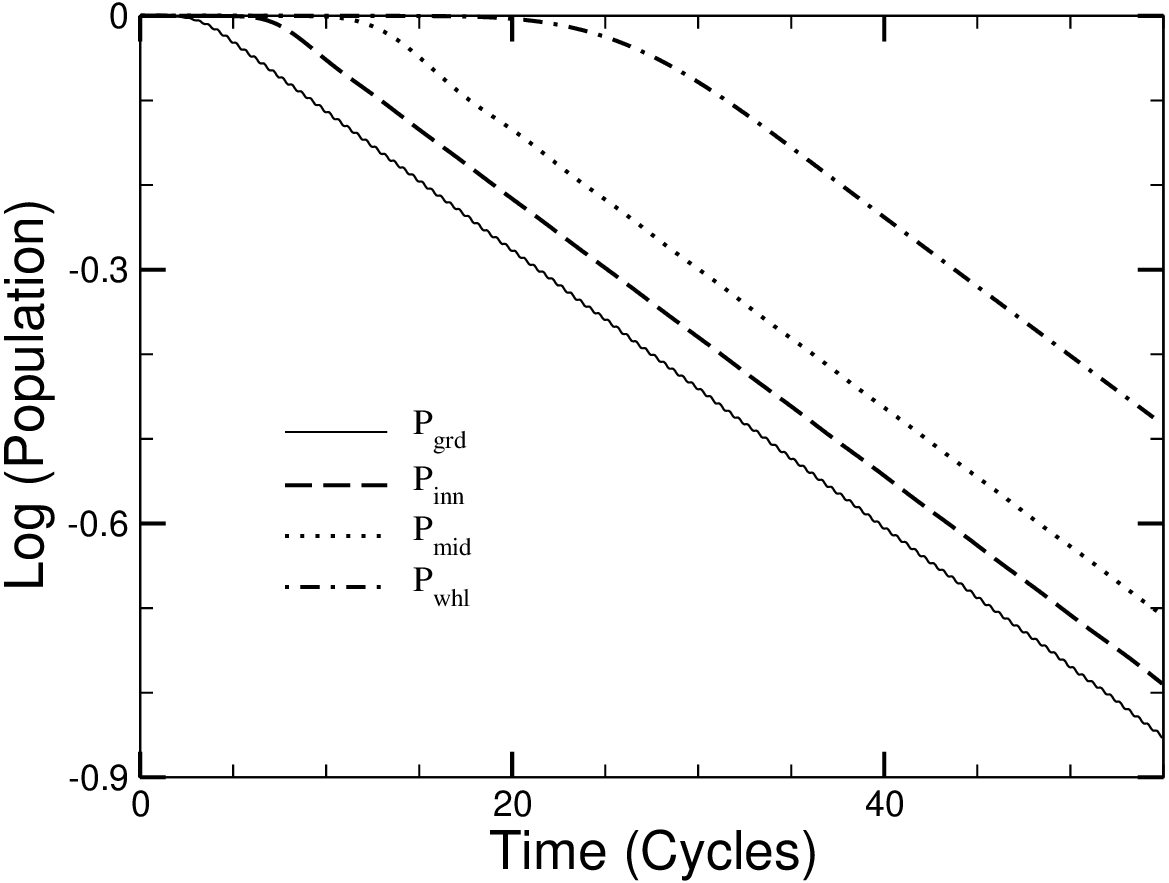}\\
    \caption{The logarithm of the decay of the ground state  and population
    within different spheres of $r$ for H at the laser frequency
    $\omega=0.6$ a.u. and the peak intensity $I_0 =
    4.375\times10^{13}$ W/cm$^2$.
     }
    \label{figure2}
\end{figure}

In figure~\ref{figure1}, we present the natural logarithm of the
population decay within a sphere of 25 a.u. of $r$ calculated by the
FD method and by the CWDVR grids corresponding to different $Z$ and
$\kappa$. With the wavelength $\lambda = 780$ nm and  peak intensity
$I_0= 2\times10^{14}$ W/cm$^{2}$, the laser pusle has a 5-cycle
ramping-on time  and  keeps constant for another 10 cycles. The
maximum number of angular momentum $L$ is taken to be 20 for a
converged calculation. In the FD case, the grid spacing $\Delta r$
is taken to be 0.2 a.u., and the number of the grid point $N=750$
for the maximum value of the grid point $r_{max}=150$. The
corresponding number of the $r$ grid points used for CWDVR cases are
indicated in table~\ref{table1}. From figure~\ref{figure1}, we
conclude that our results are indeed converged to the FD difference
results as $\kappa$ is increased from 0.5 to 2.0. Even for the
coarsest case for $\kappa=0.5$, in which case only 56 grid points
are used, the result is reasonably good. The ionization rate
estimated from this curve is 4.63$\times$10$^{13}$ s$^{-1}$, which
is close to the FD result 5.07$\times$10$^{13}$ s$^{-1}$ and the
CWDVR result 5.05$\times$10$^{13}$ s$^{-1}$ for $\kappa=2$.   We
also observe that, for the present case, results are not very
sensitive to changing of the value of $Z$(for the same value of
$\kappa$). However, we find that, in most of the cases that we
consider later, $Z=20$ is usually more preferable for its better
representation of the Coulomb potential near $r=0$.

\begin{table}[tbp]
\caption{Ionization rate $\Gamma$ (in $s^{-1}$) for ionization of H
by a linearly polarized laser of intensity $I_0$ (in W/cm$^2$) and
frequency $\omega$ (in a.u.) . The present results (a) are compared
with the previous results: (b) Chu and Cooper~\cite{Chu85}; (c)
Pont, Proulx and Shakeshaft~\cite{Pont91}; (d)
Kulander~\cite{Kula87}. The number with parenthesis like p(q) is
understood as $p\times10^{q}$. }
\begin{tabular}{|c|c|cccc|}
\hline
 $\omega$ & $I_0$ & $\Gamma^{\text a}$ & $\Gamma^{\text b}$ &
$\Gamma^{\text c}$ & $\Gamma^{\text d}$ \\ \hline
0.55 & 7.00(12) & 1.43(13) & 1.43(13) & 1.43(13) & 1.4(13) \\
 \hline
0.28 & 7.00(12) & 3.73(11) & 3.73(11) & 4.0(11) & 3.3(11) \\
& 4.38(13) & 1.33(13) & 1.33(13) & 1.35(13) & 1.2(13) \\
\hline
0.20& 4.38(13) & 3.74(12) & 3.86(12) & 4.0(12) & 2.8(12) \\
& 1.75(14) & 2.65(14) & 2.89(14) & 2.7(14) & 4.0(14) \\
& 3.94(14) & 6.14(14) & 5.65(14) & 6.0(14) & 7.0(14) \\ \hline
\end{tabular}%
\label{table2}
\end{table}

In practice, the ionization rate is fitted by an exponential decay
of the ground state population or the decay of the total population
within some sphere of $r$. In the present calculations, we have
defined an inner sphere and a middle sphere with radius $r_{inn}=25$
  and  $r_{mid}=50$ a.u. respectively.  The population remaining
  within
  these two spheres $P_{inn}$ and $P_{mid}$, together with the probability in the ground
  state  $P_{grd}$ and the total population in the whole box $P_{whl}$,
  are recorded
  for each time time step. As an example, we show in
  figure~\ref{figure2} for the ionization of H at the laser frequency
  0.6 a.u and peak intensity   $4.375\times10^{13}$ W/cm$^{2}$.
  The laser has a ramping on time of 5 cycles and keeps constant for
  50 cycles.    We observe that the population decay within different spheres of $r$
  are parallel lines with the line of the ground state decay. Therefore, it does not matter
  in this case which curve to be  used for the estimation of the ionization rate.
     The calculated ionization 1.5658$\times$10$^{-3}$ a.u. in this way  is
 in very good agreement with Chu and Cooper's result    1.5672$\times$10$^{-3}$
 a.u.~\cite{Chu85}.

  In a similar way, we have also estimated the multiphoton ionization rate of H by the
  laser of wavelength 1064 nm  at the peak intensity $1\times10^{14}$
  W/cm$^{2}$. The result is 2.97$\times$10$^{12}$ s$^{-1}$, which is in good agreement with an
  independent molecular code  result 2.85$\times$10$^{12}$ s$^{-1}$
  in cylindrical coordinates and the FD result 2.92$\times$10$^{12}$ s$^{-1}$ in spherical
  coordinates~\cite{Peng03}.

\subsection{\label{sec3sub2}Multiphoton Ionization of  H by Intense Laser Pulses}
 \begin{table}
\caption{Multiphoton ionization rates of H at different laser
electric field strengths $F_{rms}$ for different photon energies
$\omega$.  Results calculated from the present method are compared
with those of Chu and Cooper by a nonperturbative $L^2$
non-Hermitian Floquet method~\cite{Chu85}.}
\begin{ruledtabular}
\begin{tabular}{c|cccccccc}
&\multicolumn{8}{c}{$\Gamma$/2 (a.u.)}\\
\hline
 $\omega$&\multicolumn{2}{c}{$F_{rms}=0.01$
a.u.}&\multicolumn{2}{c}{$F_{rms}=0.025$ a.u.}
&\multicolumn{2}{c}{$F_{rms}=0.05$
a.u.}&\multicolumn{2}{c}{$F_{rms}=0.075$ a.u.}\\
\cline{2-3} \cline{4-5} \cline{6-7} \cline{8-9} (a.u.)
&Present&Ref~\cite{Chu85}&Present
&Ref~\cite{Chu85}&Present&Ref~\cite{Chu85}&Present
&Ref~\cite{Chu85} \\
 \hline
0.60&0.125(-3)&0.125(-3)&0.783(-3)&0.784(-3)&0.313(-2)&0.314(-2)&0.704(-2)&
0.711(-2) \\
 0.55&0.173(-3)&0.173(-3)&0.108(-2)&0.108(-2)&0.435(-2)&0.436(-2)&0.982(-2)
 &0.989(-2) \\
 0.50&0.250(-3)&0.247(-3)&0.157(-2)&0.154(-2)&0.647(-2)&0.624(-2)&0.149(-1)
 &0.139(-1) \\
 0.30&0.378(-5)&0.377(-5)&0.131(-3)&0.131(-3)&0.160(-2)&0.161(-2)&0.594(-2)
 &0.639(-2) \\
 0.28&0.451(-5)&0.451(-5)&0.161(-3)&0.161(-3)&0.204(-2)&0.204(-2)&0.821(-2)
 &0.815(-2)\\
 0.27&0.503(-5)&0.502(-5)&0.180(-3)&0.180(-3)&0.230(-2)&0.231(-2)&0.920(-2)
 &0.920(-2) \\
 0.26&0.562(-5)&0.562(-5)&0.202(-3)&0.202(-3)&0.256(-2)&0.261(-2)&0.106(-1)
 &0.110(-1) \\
 0.22&0.335(-6)&0.180(-6)&0.200(-4)&0.189(-4)&0.474(-3)&0.511(-3)&0.109(-1)
 &0.173(-1) \\
 0.20&0.140(-6)&0.106(-6)&0.452(-4)&0.467(-4)&0.321(-2)&0.350(-2)&0.743(-2)
 &0.683(-2) \\
 0.19&0.180(-4)&0.715(-5)&0.454(-3)&0.446(-3)&0.213(-2)&0.214(-2)&0.466(-2)
 &0.437(-2) \\
 0.18&0.884(-6)&0.791(-6)&0.945(-4)&0.948(-4)&0.158(-2)&0.148(-2)&0.243(-2)
 &0.376(-2)
\end{tabular}
\end{ruledtabular}
\label{table3}
\end{table}


From the above test calculations, we have seen that the present
method is applicable to different laser wave lengths at different
intensities. In this section, we give more rigorous tests of the
present method by calculating the ionization rates of H  for a large
number of wavelengths from very low to very high peak intensities.

For all the calculations presented here, we use the CWDVR grid for
$\kappa=1.0$ and $Z=20$. Therefore the number of grid point $N$ is
72 with $r_{max} =151.39 $ a.u.. The maximum number of angular
momentum $L=10$ gives converged results in all the cases for the
velocity gauge description of the interaction  term. The parameters
for the absorbing function~(\ref{MaskFun}) $\alpha$ and $\sigma$ are
taken to be 0.4 and 4 respectively.
\begin{figure}
    \includegraphics[width=8cm]{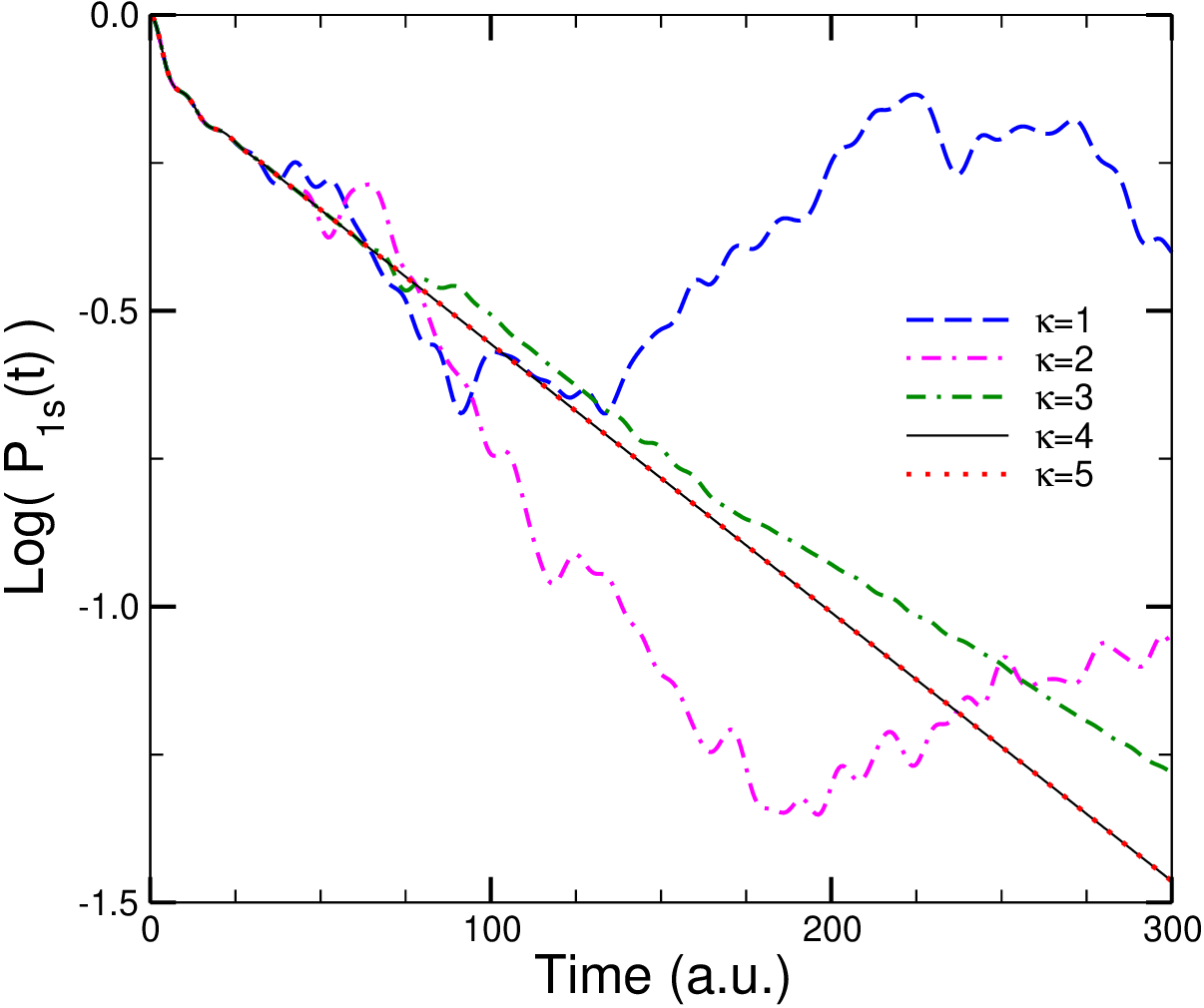}\\
    \caption{(color online). Depletion of the ground state of H by a static
    electric field with the field strength 0.08 a.u.. The ground
    state probability is shown as a function of the field duration. Results calculated
    by different CDVR grids for different $\kappa$ values are compared against the
    result calculated by FD method with $\Delta r=0.1$ a.u..}
    \label{figure3}
\end{figure}
In table~\ref{table2}, we compare ionization rates calculated using
the present method  with  other theoretical results. We notice that
our results agree better with those of Chu and Cooper~\cite{Chu85}
by an {\it ab initio} Floquet calculations. However, the present
time-dependent method results are also in reasonable agreement with
those TDSE calculations  of Ref.~\cite{Pont91} using a complex
Sturmian basis and those of  Ref.~\cite{Kula87} using a FD method.
The results of $\omega=0.2$ a.u. for intensities $1.75\times10^{14}$
and $3.94\times10^{14}$  W/cm$^{2}$ are of  less favorable agreement
because  high ionization rates lead to very fast decay of the ground
population and thus  estimations from TDSE methods become less
accurate.

In table~\ref{table3}, we compare ours results with the benchmark
calculations done by Chu and Cooper~\cite{Chu85} using the {\it ab
initio} nonperturbative $L^2$ non-Hermitian Floquet method. The
laser frequency $\omega$ vary from 0.6 to 0.18, which correspond to
one- to three-photon ionization of the ground state H.  Please note
that $E_0$ is connected with $F_{rms}$  by
 \begin{equation}
E_0 = \sqrt{2} F_{rms} = \sqrt{I/I_{au}}.
\end{equation}
Again, very good agreement are achieved for these wide ranges of
laser parameters.

\subsection{\label{sec3sub3}Ionization  of H by  Static Electric Fields}

   Now let us turn to the ionization  of H from the ground state
    by a static electric field. Excitation and ionization dynamics of atoms
    and molecules by static
electric
  fields has been of great importance since the foundation of quantum
  mechanics~\cite{Land65} and
  continues to be of great interest of current
  study~\cite{Scri00,Gelt00,Dura03}.

\begin{figure}[t]
    \includegraphics[width=12cm]{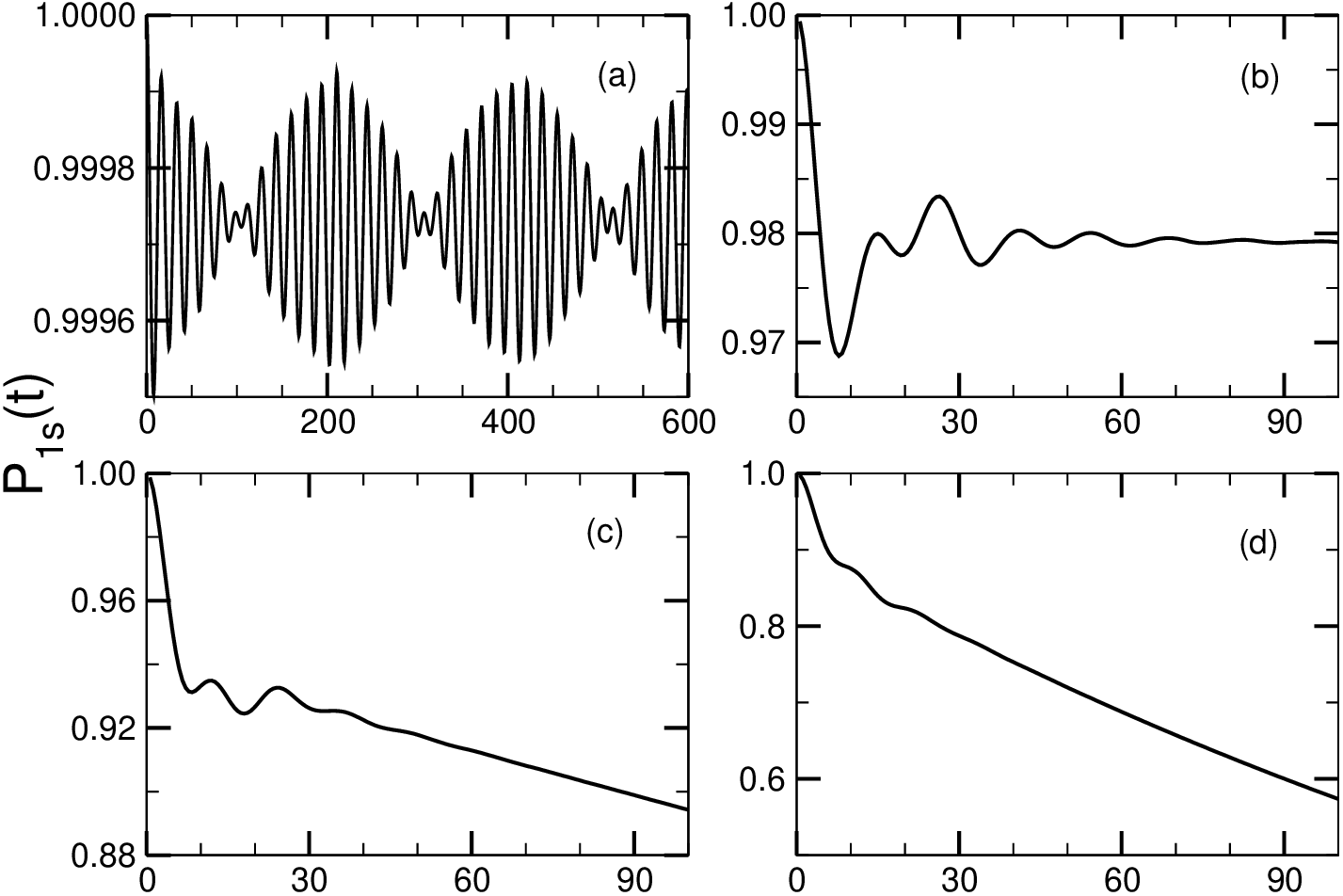}\\
    \caption{ H ground state survival probability $P_{1s}$ as a function of time
    in static electric fields. The field strength $F$ is taken
    to be: (a) 0.005; (b) 0.04; (c) 0.06; (d) 0.08 a.u.. These results are in perfect
    agreement with those by Durand and Paidarov\'a~\cite{Dura03} and by
    Scrinzi~\cite{Scri00}. }
    \label{figure4}
\end{figure}
  Although the long time behavior is dominated by the exponential decay of
  the ground state of H,   large deviations from the naive
  exponential decay are expected because of the sudden turn-on of
  static electric field. Actually, the spectrum of the Hamiltonian
  for this system
  is  unbounded from below. There are many resonances
  present.   The deviation from exponential decay is expected to  be
  large for strong fields cases~\cite{Dura03}. However, there
  is a substantially long time of transitory region for weak and
  intermediate fields as well, whose length depends on the particular
  field strength.

  In order to investigate this transitory regime, one has to make
  sure that the nonexponential  decay  indeed comes
  from physical dynamics rather than numerical reasons such as
  nonconvergence of the grid or reflection of the wave function from the edge.
  This point is extremely important for the static electric field
  case where the wave packet is driven away in one direction rather
  than in a oscillatory fashion as for a laser pulse case. In
  figure~\ref{figure3}, we show the logarithm of the ground state probability of
  H by a static electric field $F=0.08$ a.u. for different CWDVR
  grids. Note that we use the length gauge in this case for the dipole
  interaction. The maximum of the grid $r_{max}$ for all the cases
  is taken to be around 150 a.u. and the corresponding number of the
  grid points are listed in table~\ref{table1}. We take $Z=20$ for all the
  CWDVR grids.   In addition, the
  absorbing function parameters $\alpha$ and $\sigma$
  are taken to  very extreme values of 0.25 and 0.4 respectively in
  order to  avoid the reflection from the edge. For this field
  strength, we can  get converged result  only if
  $\kappa>4.0$, which case  the grid is dense enough for a good
  representation of the interaction term using the length gauge.
  Note that the performance of the present CWDVR is much better than
  the DVR method used by Dimitrovski and coworkers~\cite{Dimi03},
  in which case the converged result is only achieved for
  $t\approx40$.
  Therefore, one has be be very careful not to interpret the
  nonexponential decay for $\kappa<3.0$ as any physical excitation
  and ionization dynamics~\cite{Dura03}.

 Using the CWDVR $\kappa>4.0$ and $Z=20$, we have studied the
 short-time dynamics at other field strengths.
 In figure~\ref{figure4}, we present the population decay of the
ground state for different electric field strength $F$. For the
lowest field strength $F=0.005$ a.u., one observes both a small- and
large-time scale regular oscillation, which implies  reversible
processes. For the case when $F$ becomes 8 times of that in (a),
 we observe a quadratic decay in the first 8 a.u., followed by a
 irregular transition time before the system becomes stable after 90 a.u. or so
 (it will be a very slow exponential decay  afterwards, with $\Gamma\sim10^{-6}$ a.u.).
As the field strength increases further in (c) and (d), the
transitory time becomes shorter and shorter, which is followed by a
purely exponential decay. The transitory regions are shown  by
Durand and Paidarov\'a~\cite{Dura03} to directly related to the
$2s$-$2p$ resonances by inspecting the spectral density line-shape.
Note that, the fast quadratic  decay is present for (c) and (d) as
well when the field is turned on.

\begin{table}[t]
\caption{Ionization rate $\Gamma$ (in a.u.) for ionization of the
ground state of H by a static electric field of strength $F$ (in
a.u.). Results are compared with: (a) Scrinzi~\cite{Scri00}; (b)
Peng {\it et al}~\cite{Peng04};  (c) Bauer and Musler~\cite{Baue99}.
}
\begin{tabular}{|c|cccc|}
\hline
 $F$  & 0.06   &  0.08  & 0.1 &  0.5  \\
 \hline
 Present & 5.1509(-4) & 4.5396(-3) & 1.42(-2) &
 5.64(-1)\\
 Ref~\cite{Scri00} &5.1508(-4) & 4.5397(-3)& 1.45(-2)&5.60(-1)\\
 Others &5.15(-4)$^b$&4.55(-3)$^b$& 1.2(-2)$^c$& 5.4(-1)$^c$  \\
\hline
\end{tabular}%
\label{table4}%
\end{table}

It is remarkable to note that the present time-dependent
calculations for the entire region from very weak to very strong
field strengths are in complete agreement with the results  using
the complex scaling methods~\cite{Dura03, Scri00}.  This is further
confirmed by comparing our ionization rates fitted by an exponential
decay in the region where  the transitory time is over, which is
shown in table~\ref{table4}. Also shown in this table are the
results of some other time-dependent calculations~\cite{Baue99,
Peng04}.

\subsection{\label{sec3sub4}Multiphoton detachment rates of H$^-$ by strong laser pulse}
 \begin{table}
\caption{Multiphoton detachment rates of H$^-$ for  1064~nm, 1640~nm
and 1908~nm   at different intensities ranging from $1\times10^{10}$
to $1\times10^{12}$ W/cm$^2$. The present results from the CWDVR are
compared with those of Haritos and coworkers~\cite{Hari00} and those
of Telnov and Chu~\cite{Teln96,Teln99}. The detachment rate is
quoted as $p(q)$ which stands for $p\times 10^q$. }
\begin{ruledtabular}
\begin{tabular}{c|ccccccccc}
&\multicolumn{9}{c}{Photodetachment Rate (a.u.)}\\
\hline
 Intensity &  \multicolumn{3}{c} {1064 nm}  &  \multicolumn{3}{c} {1640 nm}  &
 \multicolumn{3}{c} {1908 nm}   \\
  \cline{2-4} \cline{5-7} \cline{8-10}
 (W/cm$^2$)&Present&Ref~\cite{Hari00}&Ref~\cite{Teln96}&Present&Ref~\cite{Hari00}&
 Ref~\cite{Teln99}&Present&Ref~\cite{Hari00}&Ref~\cite{Teln99} \\
 \hline
 $1\times10^{10}$&4.56(-5)&4.50(-5)&4.65(-5)&6.5(-7)&5.05(-7)&2.98(-7)
 &4.90(-7)&4.33(-7)&4.80(-7)\\
 $5\times10^{10}$&2.25(-4)&2.20(-4)&&7.7(-6)&6.82(-6)&&1.14(-5)&1.04(-5)&\\
 $8\times10^{10}$&3.57(-4)&3.56(-4)&&1.81(-5)&1.71(-5)&&2.81(-5)&2.58(-5)&\\
 $1\times10^{11}$&4.43(-4)&4.43(-4)&4.53(-4)&2.79(-5)&2.63(-5)&2.78(-5)&
 4.27(-5)&3.98(-5)&4.37(-5)\\
 $2\times10^{11}$&8.58(-4)&8.70(-4)&&1.01(-4)&0.99(-4)&1.05(-4)&
 1.55(-4)&1.46(-4)&1.58(-4)\\
 $3\times10^{11}$&1.25(-3)&1.28(-3)&&2.14(-4)&2.10(-4)&&3.23(-4)&3.01(-4)&\\
 $4\times10^{11}$&1.61(-3)&1.66(-3)&&3.68(-4)&3.55(-4)&3.76(-4)&5.18(-4)
 &4.95(-4)&5.30(-4)\\
 $6\times10^{11}$&2.24(-3)&2.38(-3)&&7.62(-4)&7.19(-4)&&9.52(-4)&9.78(-4)&\\
 $8\times10^{11}$&2.66(-3)&2.98(-3)&&1.16(-3)&1.15(-3)&1.23(-3)&1.48(-3)
 &1.55(-3)&1.52(-3)\\
 $9\times10^{11}$&2.81(-3)&3.24(-3)&&1.39(-3)&1.40(-3)&&1.72(-3)&1.88(-3)&\\
 $1\times10^{12}$&3.05(-3)&3.46(-3)&2.95(-3)&1.64(-3)&1.64(-3)&1.73(-3)&
 2.16(-3)&2.12(-3)&2.18(-3)
\end{tabular}
\end{ruledtabular}
\label{table5}
\end{table}

In order to show that the present method can equally apply to any
atomic systems with appropriate SAE model potential, we study the in
this section the multiphoton detachment of the negative ion of
hydrogen. For H$^-$,  we use the angular-momentum-dependent model
potential proposed by Laughlin and Chu~\cite{Laug93}
 \begin{equation}
 V_C^l(r) = \left(1+\frac{1}{r}\right)e^{-2r} -
 \frac{\alpha_d}{2r^4}W_6\left(r/r_c\right) + u_l(r),
 \end{equation}
where
 \begin{equation}
 W_j(x) = 1 - e^{-x^j},
 \end{equation}
 \begin{equation}
 u_l(r) = \left( c_0 + c_1r + c_2 r^2 \right)  e^{-\beta r}.
 \end{equation}
 In the present calculations, we use the same values of parameters
 listed in Table~I. of Ref.~\cite{Laug93}. The length gauge is used to
 describe the interaction term.
 Please also note that, we adopt the corresponding dipole operator
  which includes the
 contribution induced on the hydrogen-atom core by the outer
 loosely bounded electron.  Instead of $\mathbf{r}$, the dipole
 operator $\mathbf{D}$ in this case is given by
 \begin{equation}
 \mathbf{D} =\left[1 - \frac{\alpha_d}{r^3} W_3\left(r/r_c\right) \right]
 \mathbf{r},
 \label{dipoleHminus}
 \end{equation}

Use this model potential, the ground state we get is  $-0.027730$
a.u. for all the CWDVR grids listed in table~\ref{table1}, which is
in very good agreement with the value $-0.027733$ a.u. calculated by
Telnov and Chu~\cite{Teln04} and with the experimental measurement
$-0.027716$ a.u.~\cite{Lykk91}.

In table~\ref{table5}, we compare the total detachment rates of
H$^-$ calculated from the present method with other theoretical
calculations. We get converged results of the CWDVR grid for
$\kappa=2$ and $Z=20$. It is not surprising that the requirement of
the grid is less demanding for H$^-$ case than for the H in static
electric field case since we have a short-range potential here. For
all the laser parameters considered in table~\ref{table5}, we get
very good agreement with those results of Refs.~\cite{Hari00}
and~\cite{Teln96, Teln99}.

However, Telnov and Chu use the dipole moment $\mathbf{r}$, which
accounts for the slight differences between their results and ours.
We have tested our code with the latter dipole moment, we get
detachment rate to be $4.51\times 10^{-4}$ a.u. instead of
$4.43\times 10^{-4}$ a.u. for 1064~nm at $1\times10^{11}$ W/cm$^2$.
 On the other hand, our results are slightly closer to those of
 Haritos and coworkers~\cite{Hari00}, which are calculated by
 solving the time-independent Schr\"odinger equation by the
 nonperturbative many-electron, many-photon theory (MEMPT). This
 might be due to the fact that the dipole moment we use takes the
 core polarization effects into account.

\section{\label{sec4}Conclusions}

  We have presented an accurate and efficient method of solving the
  time-dependent Schr\"odinger equation. Compared to the usual
  FD discretization scheme of the radial coordinate,
  the present CWDVR method needs 3-10 fewer number of grid points
  and treats the Coulomb singularity naturally and accurately. As
  examples, our method has been very successfully applied to the
  multiphoton ionization dynamics of both H and H$^-$.
  The total ionization rates estimated from the present method are
  in perfect agreement with other theoretical results. We have also
  applied
  our method to investigate the short-time excitation and ionization dynamics
  of H in weak and strong static electric fields. The ground state survival
  probability and the ionization rate calculated using the present method are in
   full agreement with those results calculated by complex
  rotation method.  Since the CWDVR treats the Coulomb potential
  accurately and needs much fewer points in $r$ coordinate, the
  present methods opens the way to a more efficient treatment of
  many-electron atomic systems.

\begin{acknowledgments}
L.Y.P.  appreciates helpful discussions with Profs. D.A. Telnov and
S.I. Chu about gauge choice of H$^-$ and communications with Prof.
R.B. Sidje about application of the package ``Expokit''. This
research is partially supported by NSF and DOE.
\end{acknowledgments}

\bibliographystyle{AABBRV}
\bibliography{ACOMPAT,harvard}

\end{document}